\documentclass[12pt]{article}
\usepackage{times}
\usepackage{geometry}
\usepackage[utf8]{inputenc}
\usepackage{enumitem,amssymb}
\usepackage{ragged2e}
\newlist{thematic}{itemize}{8}
\setlist[thematic]{label=$\square$}
\usepackage{pifont}
\usepackage{graphicx}

\begin{document}
\raggedright
\huge
Astro2020 Science White Paper \linebreak

Intermediate-Mass Black Holes
\medskip
in Extragalactic Globular Clusters \linebreak
\normalsize

\noindent \textbf{Thematic Areas:} \hspace*{60pt} $\square$ Planetary Systems \hspace*{10pt} $\square$ Star and Planet Formation \hspace*{20pt}\linebreak
$\boxtimes$ Formation and Evolution of Compact Objects \hspace*{31pt} $\square$ Cosmology and Fundamental Physics \linebreak
  $\square$  Stars and Stellar Evolution \hspace*{1pt} $\square$ Resolved Stellar Populations and their Environments \hspace*{40pt} \linebreak
  $\boxtimes$    Galaxy Evolution   \hspace*{45pt} $\boxtimes$             Multi-Messenger Astronomy and Astrophysics \hspace*{65pt} \linebreak
  
\textbf{Principal Author:}

Name:	Joan M. Wrobel
 \linebreak						
Institution:  National Radio Astronomy Observatory, USA
 \linebreak
Email: jwrobel@nrao.edu
 \linebreak
Phone:  1-575-418-7511
 \linebreak
 
\textbf{Co-authors:} (names and institutions)
 \linebreak
Zoltan Haiman, Columbia University, USA
 \linebreak
Kelly Holley-Bockelmann, Vanderbilt University, USA
 \linebreak
Kohei Inayoshi, Peking University, China
 \linebreak
Joe Lazio, Jet Propulsion Laboratory, Caltech, USA
 \linebreak
Tom Maccarone, Texas Tech University, USA 
 \linebreak
James Miller-Jones, Curtin University, Australia
 \linebreak
Kristina Nyland, National Research Council, resident at the Naval Research Lab, USA 
 \linebreak
Rich Plotkin, University of Nevada, USA
 \linebreak

\textbf{Abstract (optional):} Intermediate-mass black holes (IMBHs)
have masses of about 100 to 100,000 solar masses.  They remain elusive.  
Observing IMBHs in present-day globular clusters (GCs) would validate a
formation channel for seed black holes in the early universe and inform 
event predictions for gravitational wave facilities.  Reaching a large
number of GCs per galaxy is key, as models predict that only a few 
percent will have retained their gravitational-wave fostering IMBHs.  
Related, many galaxies will need to be examined to establish a robust 
sample of IMBHs in GCs.  These needs can be meet by using a 
next-generation Very Large Array (ngVLA) to search for IMBHs in the GCs
of hundreds of galaxies out to a distance of 25 Mpc.  These galaxies 
hold tens of thousands of GCs in total.  We describe how to convert an 
ngVLA signal from a GC to an IMBH mass according to a semi-empirical 
accretion model.  Simulations of gas flows in GCs would help to improve
the robustness of the conversion.  Also, self-consistent dynamical models 
of GCs, with stellar and binary evolution in the presence of IMBHs, 
would help to improve IMBH retention predictions for present-day GCs.
\pagebreak

\textbf{1. Globular Star Clusters as Hosts of Intermediate-Mass Black Holes} 
\linebreak 
\linebreak 
Both theory and computational modeling suggest that globular clusters 
(GCs) can host intermediate-mass black holes (IMBHs) with masses $M_{IMBH} 
\sim 100-100,000~M_\odot$, but they have yet to be convincingly found in
such settings [reviewed in 27].  Detecting IMBHs in GCs would validate one
formation channel for seed black holes (BHs) in the early universe 
[reviewed in 22] and have broad implications for the GCs' dynamical 
evolution [19].  GCs reside in galaxy halos, with each galaxy hosting a 
system of a hundred to a thousand GCs [17].  Studying these so-called GC 
systems could constrain the ability of GCs to retain IMBHs and provide key
input into event predictions for gravitational wave (GW) facilities.  For 
example, a GC system with a low fraction of IMBHs at present could be 
linked to a high rate of GW events in the past [13,14,19].  Observing a 
large number of GCs is critical, as [13,14,19] predict that only a few 
percent will have retained their IMBHs.  This implies that to gain a
robust picture of IMBHs in GCs, many galaxies must be examined and each 
galaxy's GC system must be well sampled.  Figure~1 shows the distances of
galaxies whose GC systems have been studied [17], with an inset displaying 
an example GC system [8].  The effective diameter of a GC system is a few
tens of kpcs [12] and the half-starlight diameter of a typical GC itself 
is 5~pc [7]. \linebreak 

\begin{figure}[!htb]
\centering\includegraphics[scale=0.5]{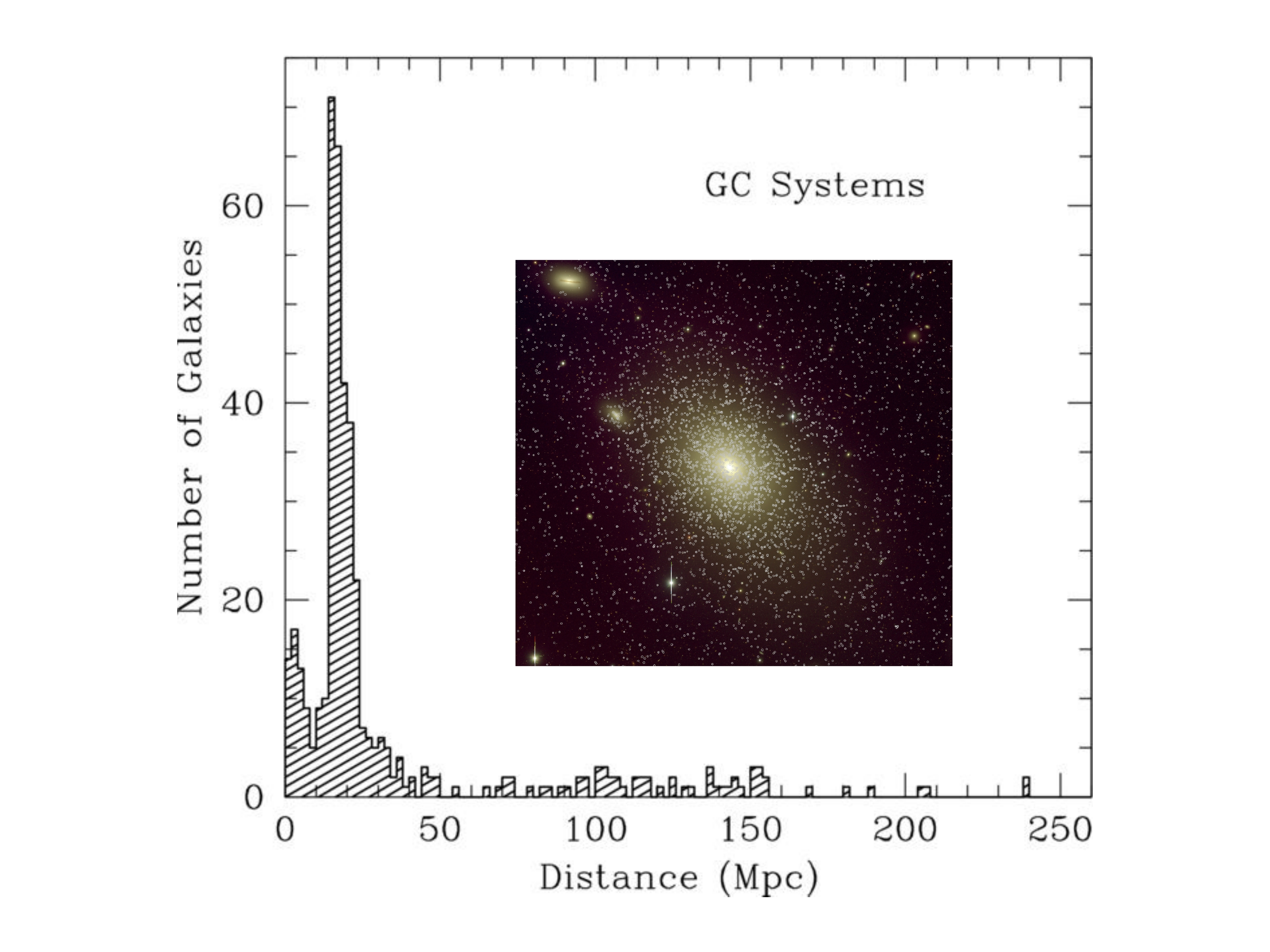}
\vspace{-20pt}
\caption{\em Distances of galaxies whose systems of GCs have been studied 
  [17].  The inset shows an example GC system, of the early-type galaxy NGC\,4365 
  at 23~Mpc [8].  The sprinkling of symbols mark GC candidates in the 
  inner 18$'$ (120 kpc) of a {\em gri} Suprime-Cam image.}\label{f1}
\end{figure}

To search for IMBHs in GCs, one looks for evidence that the IMBHs are
affecting the properties of their GC hosts [reviewed in 27, 35].  In
the Local Group, a common approach is to use optical or infrared data
to look for the dynamical signatures of IMBHs on the orbits of stars
in the GCs.  Such sphere-of-influence studies have a contentious history
[5], even leading to differing IMBH masses when using the orbits of stars
or of radio pulsars in the same GC [16,30].  A basic limitation of 
dynamical searches is that they are susceptible to measuring high 
concentrations of stellar remnants rather than an IMBH [26,35].  To make 
progress, it is important to develop independent approaches that bypass 
these issues. 
\linebreak 
\linebreak 
One promising approach is to search for radio signatures of accretion 
from IMBHs in GCs [reviewed in 25].  This approach leverages on decades 
of studies of the signatures of accretion onto stellar-mass and 
supermassive BHs [reviewed in 11].  Section~2 gives a synopsis of the 
synchrotron radio model.  Section~3 describes applying the model to two 
galaxies and underscores the sensitivity shortfalls of radio facilities 
into the 2020s.  Section~4 examines the prospects for applying the model 
to many galaxies in the 2030s.  Section~5 describes ways to improve the 
model's theoretical underpinnings and interpretive framework.
\linebreak \linebreak

\textbf{2. Synchrotron Radio Model} 
\linebreak 
\linebreak 
Following [25,34], we invoke a semi-empirical model to predict the mass
of an IMBH that, if accreting slowly from the tenuous gas supplied by 
evolving stars, is consistent with the synchrotron radio luminosity of a
GC.  We assume gas-capture at 3\% of the Bondi rate [29] for gas at a 
density of 0.2 particles~cm$^{-3}$ [1] and at a constant temperature of 
10,000~K.  We also assume that accretion proceeds at less than 2\% of the
Eddington rate, thus involving an inner advection-dominated accretion 
flow with a predictable, persistent X-ray luminosity.  (An IMBH accreting 
at higher than 2\% of the Eddington rate would enter an X-ray-luminous 
state [24] and be easily detectable in existing surveys.  But no such 
X-ray sources exist in Milky Way GCs.)  We then use the empirical 
fundamental-plane of BH activity as refined by [32] to predict the 
synchrotron radio luminosity.  The radio emission is expected to be 
persistent, flat-spectrum, jet-like but spatially unresolved, and located 
at or near the dynamical center of the GC.  \linebreak \linebreak

\textbf{3. Observational Shortfalls into the 2020s}
\linebreak 
\linebreak
We have used the NSF's Karl G.\ Jansky Very Large Array (VLA) [31] at 
a 6-cm wavelength and 20-pc resolution to search for the signatures of 
accretion from IMBHs in 337 candidate GCs in NGC\,1023 [36] and 206 
probable GCs in M81 [37].  None of the individual GCs were detected.  
From the radio synchrotron model, the lowest mass limits inferred were 
$M_{IMBH} < 390,000~M_\odot$ for NGC\,1023 at 11~Mpc and 
$M_{IMBH} < 100,000~M_\odot$ for M81 at 3.6~Mpc.  Our stacking analysis 
of each GC system achieved about a factor of two improvement in the IMBH 
mass limits.  The stacks assumed that each GC system had a high IMBH 
retention fraction and a uniform IMBH mass distribution.  So far, 
modelling suggests that neither assumption is likely to be valid 
[13,14,19], which weakens any inferences from the VLA stacks.
\linebreak
\linebreak
For a dozen GCs in M81, the upper limits on their IMBH-to-stellar mass 
ratios were less than 0.15 [37].  Mass ratios in that regime are observed
in some ultracompact dwarf galaxies [2] and predicted for some 
present-day GCs [14].  Thus, our VLA study of M81 provides a first
glimpse of the potential for deeper radio searches to constrain IMBHs in 
GCs out to tens of Mpcs.
\linebreak 
\linebreak
In the 2020s the deployment baseline of SKA1-Mid plans to offer a 
spatial resolution of 57~mas at wavelength of 3~cm [6,9]. 
Declinations south of +10 degrees will be viewable.  At a distance of
25~Mpc the spatial resolution corresponds to 7~pc, close to a GC's
half-starlight diameter of 5~pc [7].  Only 67 SKA1-Mid antennas will
be available at 3~cm.  This means that the effective collecting area of 
SKA1-Mid will be about that of the current VLA, which is insufficient to
reach many galaxies.
\linebreak 
\linebreak 
\textbf{4. Observational Imperatives in the 2030s} 
\linebreak 
\linebreak
\textbf{4.1. Applying the Synchrotron Radio Model in the 2030s} 
\linebreak
\linebreak 
In [38], we considered using Band 3 of the next-generation Very Large
Array (ngVLA) [33] to examine GC systems out to distances of tens of
Mpcs.  Band 3 has a central frequency of 17~GHz and a bandwidth of
8.4~GHz.  Its central wavelength is 2~cm.  The field of view (FOV) is
a circle of diameter 3.6$'$ at full width half maximum.  Declinations
north of -40 degrees will be viewable.
\linebreak 
\linebreak 
The compilation of GC systems presented in Figure~1 involves 422 
galaxies [17].  The distribution of the galaxies' distances shows two 
peaks.  A minor peak contains tens of galaxies, either isolated or group
members, with distances out to 10~Mpc.  A major peak contains hundreds
of galaxies with distances between 10 and 25~Mpc, including members of
the Virgo and Fornax clusters of galaxies.  The major peak in distance
contains tens of thousands of GCs in total.  These GC counts can be 
further boosted by the recently-recognized population of intracluster 
GCs in the Virgo Cluster [23].
\linebreak 
\linebreak 
In [38], we applied the synchrotron model to predict the luminosity at
2~cm as a function of the mass, $M_{IMBH}$, of a putative IMBH in a GC.
We then derived the associated point-source flux densities, $S_{2cm}$,
for GCs at distances of 10 and 25~Mpc.  In Figure~2, the sloping lines
show how to convert from $S_{2cm}$ to $M_{IMBH}$ for the two distances,
while the vertical lines show 3$\sigma$ detections with the ngVLA,
assuming tapered and robust weighting, with integrations of 1, 10 and
100~hours [33].  At higher signal-to-noise ratios, the wide frequency
coverage could test the flat-spectrum prediction, as well as raise flags
about steep-spectrum contaminants.  (Potential radio contaminants in
extragalactic GCs were considered and ruled out by [38].)
\linebreak 
\linebreak 
From Figure~2, the synchrotron model predicts a flux density of
$S_{2cm} = 0.27~\mu$Jy from an IMBH of mass 76,000~$M_\odot$ at 10~Mpc
or of mass 150,000~$M_\odot$ at 25~Mpc.  The main subarray of the
ngVLA can make 3$\sigma$ detections with 10 hours on target and a
tapered, robustly-weighted resolution of 100~mas [33].  This spatial
resolution matches the half-starlight diameter of 5~pc [7] for a
GC at 10~Mpc and suffices to localize the source to a GC at 25~Mpc.
A GC system has an effective diameter of a few tens of kpcs [12], so
at these distances most of it can be encompassed in a few FOVs.  An
important consequence is that each 3.6$'$ FOV can simultaneously 
capture many GCs.  To summarize, with its sensitivity, bandwidth, spatial
resolution, and FOV, the ngVLA at a wavelength of 2~cm could efficiently 
probe IMBH masses in hundreds of GC systems out to a distance of 25~Mpc. \textbf{\em To 
search for these IMBHs, we recommend constructing the next-generation Very Large Array (ngVLA).}
\linebreak 

\begin{figure}[!htb]
\centering\includegraphics[scale=0.6]{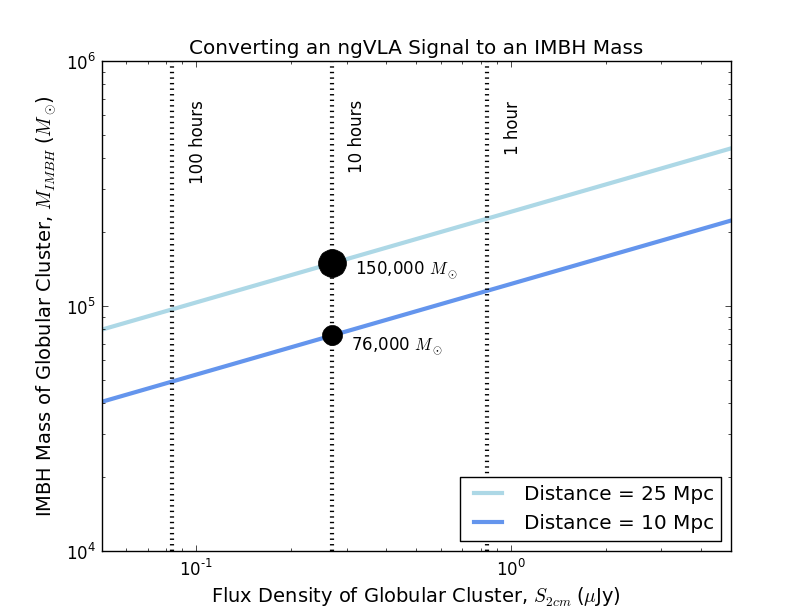}
\vspace{-10pt}
\caption{\em ngVLA signals, $S_{2cm}$, from IMBH masses, $M_{IMBH}$,
  in GCs at distances of 10 and 25~Mpc.  The small and big
  black dots highlight 3$\sigma$ mass sensitivities at 10 and 25~Mpc,
  respectively, after 10 hours on target.  Adapted from [38].}\label{f2}
\end{figure}

\textbf{4.2 Radio Synergies with Gravitational Wave Facilities in the 2030s}
\linebreak 
\linebreak 
The fate of primordial GCs, each born with a central IMBH, was explored
by [13,14,19].  They modelled the evolution of the GCs in a variety of 
host galaxies, and of the IMBHs undergoing successive, GW-producing 
mergers with stellar-mass BHs in the GCs.  For primordial GCs that 
survived to the present day, only a few percent retained their IMBHs and
the balance lost their IMBH when a GW recoil ejected it from the GC host.
Once ejected, the IMBHs are no longer able to foster GW events.  Retained 
IMBHs gained mass via tidal disruption events and captures of 
stellar-mass BHs, while host GCs lost mass due to stellar winds and 
tidal stripping [14].  Illustrative present-day GCs achieved high values
for IMBH masses ($M_{IMBH} \sim 160,000~M_\odot$) and IMBH-to-stellar mass 
ratios (6-12\%).
\linebreak
\linebreak
These results were used to predict volumetric event rates for planned GW 
facilities ({\em LISA} [4], Einstein Telescope [18], Cosmic Explorer [3]).
IMBHs with masses between 1000 and 10,000~$M_\odot$ yielded mergers at 
rates that could be detected by all three GW facilities.  IMBHs with masses
$\gtrsim 10,000~M_\odot$ yielded mergers at rates that could be detected 
only by {\em LISA}, slated to begin observing in 2036 or earlier [4].  
{\em LISA} would also be able to detect stellar-mass BHs and stellar 
remnants as they interact with the IMBH [4], which could provide important 
constraints on the stellar population and dynamical state of the host GC.
In the interim, ngVLA searches for IMBHs in present-day GCs could occur, 
via Early Science notionally starting in 2028 or via Full Science notionally 
starting in 2034.  If the ngVLA searches do not match the present-day 
predictions, it would challenge the framework underlying the merger-rate 
predictions for all three planned GW facilities.
\linebreak 
\linebreak
\textbf{4.3 Radio Synergies with Other Electromagnetic-Wave Facilities in the 2030s}
\linebreak 
\linebreak
A key science driver for extremely large telescopes (ELTs) in the 30-m
class is to measure, at a distance of 10~Mpc, a BH mass as low as
$M_{IMBH} \sim 100,000~M_\odot$ by spatially resolving its sphere of
influence in its GC host [10].  For example, if the Infrared Imaging
Spectrometer (IRIS) on the Thirty Meter Telescope (TMT) can achieve 
the diffraction limit of 18~mas at 2~$\mu$m, then this approach could
yield a sample of IMBHs in GCs out to a distance of 10~Mpc.  An IRIS
study must be done one GC at a time, a shortcoming that makes it
expensive to inventory many GCs per galaxy. The TMT's Infrared
Multi-object Spectrometer (IRMS) in spectroscopy mode will have a FOV
of 2.0$'$ $\times$ 0.6$'$.  This being a tenth of the ngVLA FOV, a
sphere-of-influence study with IRMS would require ten pointings to 
cover one ngVLA pointing.  Regardless of the situation out to 10~Mpc, 
the ELT approach cannot reach the hundreds of GC systems with 
distances between 10 and 25~Mpc.  A white paper by Greene et al. covers further aspects of IMBH searches using ELTs.
\linebreak 
\linebreak
The {\em Chandra} X-ray mission and its proposed successors, {\em
  Lynx} [15] and the {\em Advanced X-ray Imaging Satellite} [28], feature
spatial resolutions of 300 to 500~mas.  These will suffice to roughly
localize X-ray sources to GCs out to a distance of 25~Mpc.  But an
X-ray--only search for the accretion signatures of IMBHs in GCs will
be hindered by confusion from X-ray binaries in GCs [21].
Specifically, X-ray--only detections of GCs cannot discriminate
between X-ray binaries and IMBHs.  Fortunately, the empirical
fundamental-plane of BH activity as refined by [32] implies that
the persistent radio emission from IMBHs is expected to be several
hundred times greater than that from X-ray binaries.  Thus ngVLA
imaging can be used to separate X-ray detections into bins for X-ray
binaries and for IMBHs.  X-ray binaries are known to be time-variable
in both the radio and X-ray bands, so this radio--X-ray synergy would
be strengthened by simultaneous observations with the ngVLA and the
X-ray mission.  White papers by Gallo et al.\ and Haiman et al.\ cover further aspects of IMBH searches using X-ray facilities.
\linebreak 
\linebreak
\textbf{5. Theoretical Imperatives in the 2030s}
\linebreak 
\linebreak
Self-consistent dynamical models of GCs, with stellar and binary 
evolution in the presence of IMBHs, would help to improve IMBH retention
predictions for present-day GCs, and should be pursued with priority.
If these models yield higher IMBH retention fractions, one could stack
the ngVLA data for a galaxy's GC system and improve IMBH mass 
sensitivities by about a factor of two.  Also, from parameter 
uncertainties, [34] estimate that the IMBH mass associated with a 
given radio luminosity could be in error by a factor of 2.5.  To 
improve the robustness of such masses, the assumed gas-capture rate 
from a Bondi flow should be replaced by realistic simulations 
[e.g., 20] of gas flows in GCs.
\linebreak 
\linebreak
\textbf{Acknowledgement:} The NRAO is a facility of the NSF, operated
under cooperative agreement by Associated Universities, Inc.

\pagebreak
\textbf{References}

[1] Abbate, F., et al.\ 2018, MNRAS, 481, 627

[2] Ahn, C.~P., et al.\ 2017, ApJ, 839, 72

[3] Abbott, B.~P., et al.\ 2016, arXiv:1607.08697

[4] Amaro-Seoane, P., et al.\ 2017, arXiv:1702.00786

[5] Baumgardt, H.\ 2017, MNRAS, 464, 2174

[6] Borjesson, L.\ 2017, SKA Board Meeting, 2017, July 18-19

[7] Brodie, J.~P., \& Strader, J.\ 2006, ARAA, 44, 193

[8] Brodie, J.~P., et al.\ 2014, ApJ, 796, 52

[9] Dewdney, P., et al.\ 2015, SKA-TEL-SKO-0000308

[10] Do, T., et al.\ 2014, AJ, 147, 93

[11] Fender, R., \& Munoz-Darias, T.\ 2016, in Astrophysical Black
Holes, eds. F.\ Haardt et al.\ (Springer), 65

[12] Forbes, D.~A. 2018, MNRAS, 472, L104

[13] Fragione, G., et al.\ 2018a, ApJ, 856, 92

[14] Fragione, G., et al.\ 2018b, ApJ, 867, 119

[15] Gaskin, J.~A., et al. \ 2018, SPIE, 106990N

[16] Gieles, M., et al.\ 2017, MNRAS, 473, 4832

[17] Harris, W.~E., et al.\ 2013, ApJ, 772, 82

[18] Hild, S., et al.\ 2011, Class.\ Quantum Grav., 2011, 28, 094813

[19] Holley-Bockelmann, K., et al.\ 2008, ApJ, 686, 829

[20] Inayoshi, K., et al.\ 2018, MNRAS, 476, 1412

[21] Joseph, T.~D., et al.\ 2017, MNRAS, 470, 4133

[22] Katz, K. 2018, arXiv:1807.06593

[23] Longobardi, A., et al.\ 2018, ApJ, 864, 36

[24] Maccarone, T.~J.\ 2003, A\&A, 409, 697

[25] Maccarone, T.~J.\ 2016, Mem.\ S.\ A.\ It., 87, 559

[26] Mann, C.~R., et al.\ 2018, arXiv:1807.03307

[27] Mezcua, M.\ 2017, Int.\ J.\ Mod.\ Phys.\ D, 26, 1730021

[28] Mushotzky, R., et al.\ 2018, arXiv:1807.02122

[29] Pellegrini, S.\ 2005, ApJ, 624, 155

[30] Perera, B.~B.~P., et al.\ 2017, MNRAS, 468, 2114

[31] Perley, R.~A., et al.\ 2011, ApJL, 739, L1

[32] Plotkin, R.~M., et al.\ 2012, MNRAS, 419, 267

[33] Selina, R.~J., et al.\ 2018, in Science with a Next Generation
Very Large Array, ed. E.~J. Murphy (ASP), 15

[34] Strader, J., et al.\ 2012, ApJL, 750, L27

[35] van der Marel, R.\ 2013, SnowPAC 2013 - Black Hole Fingerprints

[36] Wrobel, J.~M., et al.\ 2015, AJ, 150, 120

[37] Wrobel, J.~M., et al.\ 2016, AJ, 152, 22

[38] Wrobel, J.~M., et al.\ 2018, in Science with a Next Generation
Very Large Array, ed. E.~J. Murphy (ASP), 743

\end{document}